\documentclass[RNAAS]{aastex631}

\usepackage{graphicx}
\usepackage{color}
\usepackage{amsbsy}
\usepackage{mathrsfs}
\usepackage{mathpazo,bm}
\usepackage{amsmath}
\usepackage{xfrac}
\usepackage{url}
\usepackage{hyperref}
\usepackage{enumitem}
\usepackage{natbib}
\usepackage{amssymb}
\usepackage{enumitem}
\usepackage{ulem}
\usepackage{wrapfig}

\newcommand{\beq}{\begin{equation}}
\newcommand{\eeq}{\end{equation}}
\newcommand{\beqn}{\begin{eqnarray}}
\newcommand{\eeqn}{\end{eqnarray}}
\newcommand{\pderiv}[2]{\frac{\partial{#1}}{\partial{#2}}}

\renewcommand{\v}[1]{{\boldsymbol{#1}}} 

\newcommand{\del}{\v{\nabla}}
\newcommand{\grad}{\del}
\newcommand{\Div}{\del\cdot}

\newcommand{\Laplace}{\nabla^2}

\newcommand{\haty}{\hat{\v{y}}}
\newcommand{\hatx}{\hat{\v{x}}}

\newcommand{\Eq}[1]{Eq.~(\ref{#1})}

\newcommand{\eq}[1]{\Eq{#1}}

\definecolor{brown}{rgb}{0.42,0.24,0.07}
\definecolor{darkgreen}{rgb}{0.0,0.6,0.00}
\definecolor{purple}{rgb}{0.7,0.0,0.7}
\definecolor{black}{rgb}{0.0,0.0,0.0}

\def\white#1{\textcolor{white}{#1}}

\def\apj{\rm ApJ}
\def\apjl{\rm ApJL}
\def\apjs{\rm ApJS}

\def\aap{\rm A\&A}

\newcommand{\eps}{\epsilon}

\newcommand{\hmu}{{\hat{\mu}}}
\newcommand{\hnu}{{\hat{\nu}}}
\newcommand{\he}{{\hat{e}}}
\newcommand{\hatm}{{\hat{m}}}
\newcommand{\hatn}{{\hat{n}}}

\shorttitle{Vortex in ellitpic coordinates}
\shortauthors{Lyra}

\begin{document}

\title{Vortex solution in elliptic coordinates}

\author[0000-0002-3768-7542]{Wladimir Lyra}
\affiliation{New Mexico State University, Department of Astronomy, PO
  Box 30001 MSC 4500, Las Cruces, NM 88001, USA}







\begin{abstract}

Vortices (flows with closed elliptic streamlines) are exact nonlinear solutions to the compressible Euler
equation. In this contribution, we use differential geometry to derive the
transformations between Cartesian and elliptic coordinates,
and show that in elliptic coordinates a constant vorticity flow
reduces to $\dot{\mu}=0$ and $\dot{\nu}={\rm
  const}$ along the streamline $\mu_0$ that matches the vortex eccentricity. 
\end{abstract}



\section{Introduction} \label{sec:intro}

Vortices are important for planet formation, theorized as favorable
locations for dust trapping \citep{BargeSommeria95}. Crescent-shaped asymmetries have
been observed in sub-mm images of protoplanetary disks
\citep{vanderMarel+13}, although their unambiguous
identification vortices has been elusive. A patch of constant
vorticity follows the solution $\v{u} = \Omega y/\chi \hatx,
-x\chi \haty$, where $(x,y)$ are the Cartesian coordinates, $\Omega$ is a
constant and $\chi=x/y>1$ is the vortex aspect ratio. Given the
elliptic streamlines, a solution in terms of elliptic coordinates $(\mu,\nu)$ is desirable.

\section{Elliptical coordinates}

The orthogonal elliptical coordinate system is 

\beqn
  x &=& f\cosh\mu\cos\nu, \\
  y &=& f\sinh\mu\sin\nu,
\eeqn

\noindent where $f=a\eps$ is the focal length, $a$ the semi-major 
axis, and $\eps$ the eccentricity. Constant 
$\mu$ define ellipses, constant $\nu$ define hyperbolae.
The coordinates describe confocal ellipses: the focal distance is 
constant, so changing $\mu$ changes not only the 
semimajor axis but also the eccentricity. 

\subsection{Metric}

The metric of this system is 

\beqn
  g_{ij} &=& \pderiv{x^\alpha}{q^i}\pderiv{x^\beta}{q^j} g_{\alpha\beta},\nonumber\\
        &=& f^2\left(\sinh^2\mu + \sin^2\nu\right)\,\delta_{ij}.
\eeqn

\noindent where $\v{x}=(x,y)$ and $\v{q}=(\mu,\nu)$ are Cartesian and
elliptic coordinates; $g_{\alpha\beta} = \delta_{\alpha\beta}$ is the metric of
Cartesian space. From this transformation, it follows that the scale factors are equal 

\beqn
  h_\mu &=& \sqrt{g_{\mu\mu}} = f\sqrt{\sinh^2\mu + \sin^2\nu},\\
  h_\nu &=& \sqrt{g_{\nu\nu}} = f\sqrt{\sinh^2\mu + \sin^2\nu}.
\eeqn

We hereafter use $h=h_\mu=h_\nu$. We also use the equivalent definition

\beq
h  = \frac{f}{\sqrt{2}}\sqrt{\cosh\,2\mu - \cos\,2\nu}.
\label{eq:otherh}
\eeq

The derivatives with respect to the coordinates are 

\beqn
  \partial_\mu h &=& \frac{f^2}{2h} \, \sinh\,2\mu, \\
  \partial_\nu h &=& \frac{f^2}{2h} \, \sin\,2\nu.
\eeqn

We calculate the Christoffel symbols in non-coordinate basis

\beqn
  \Gamma_{\hat\alpha\hat\beta\hat\gamma} &=& \frac{1}{2}\left(c_{\hat\alpha\hat\beta\hat\gamma} + c_{\hat\alpha\hat\gamma\hat\beta} - c_{\hat\beta\hat\gamma\hat\alpha}\right),\\
  \Gamma^{\hat\alpha}_{\hat\beta\hat\gamma} &=& g^{\hat\alpha\hat\zeta}\Gamma_{\hat\zeta\hat\beta\hat\gamma},
\eeqn

\noindent where $c_{\hat\beta\hat\gamma\hat\alpha}=g_{\hat\alpha\hat\zeta}{c_{\hat\beta\hat\gamma}}^{\hat\zeta}$ are the
connection coefficients, given by 

\beq
  [e_{\hat\beta},e_{\hat\gamma}] = {c_{\hat\beta\hat\gamma}}^{\hat\alpha} \partial_{\hat\alpha}.
\eeq

Given that $e_\hmu = h^{-1}\partial_\mu$ and $e_\hnu = h^{-1}\partial_\nu$, we have 

\beqn
  [e_\hmu,e_\hnu] &=& \frac{1}{h}\left[\pderiv{}{\mu}\left(\frac{1}{h}\pderiv{}{\nu}\right) - 
                                  \pderiv{}{\nu}\left(\frac{1}{h}\pderiv{}{\mu}\right)\right]\nonumber\\
               &=&\frac{f^2}{2h^4} \left(\sin\,2\nu\pderiv{}{\mu} - \sinh\,2\mu\pderiv{}{\nu}\right) \nonumber\\
               &=&\frac{f^2}{2h^3} \left(\sin\,2\nu\,\partial_\hmu - \sinh\,2\mu\,\partial_\hnu\right) \nonumber\\
               &=&-[e_\hnu,e_\hmu].
\eeqn

The connection coefficients are thus

\beqn
  {c_{\hmu\hnu}}^\hmu = -{c_{\hnu\hmu}}^\hmu &=& \white{-}\frac{f^2}{2h^3}\sin\,2\nu,\\  
  {c_{\hmu\hnu}}^\hnu = -{c_{\hnu\hmu}}^\hnu &=& -\frac{f^2}{2h^3}\sinh\,2\mu.
\eeqn

\noindent And the Christoffel symbols are 

\beqn
 \Gamma^\hmu_{\hnu\hmu} = -\Gamma^\hnu_{\hmu\hmu} &=& \frac{f^2}{2h^3}\sin\,2\nu,\\  
 \Gamma^\hnu_{\hmu\hnu} = -\Gamma^\hmu_{\hnu\hnu} &=& \frac{f^2}{2h^3}\sinh\,2\mu.
\eeqn

The elliptic and Cartesian unit vectors $\hat{e}_i$ and $\hat{x}_i$
transform according to 

\beq
  \hat{e}_i = \frac{1}{h_i}\frac{\partial x_j}{\partial e_i} \hat{x_j}, 
\eeq

\noindent i.e.,

\beqn
   \left[\begin{array}{c}
       \hat\mu \\
       \hat\nu \\
     \end{array}\right]&=&\left[\begin{array}{ccc}
       h_\mu^{-1}\partial_\mu x &\white{..} & h_\mu^{-1}\partial_\mu y\\
       h_\nu^{-1}\partial_\nu x &\white{..} & h_\nu^{-1}\partial_\nu y\\
     \end{array}\right]\left[\begin{array}{c}
       \hat{x} \\
       \hat{y}\\
     \end{array}\right] \nonumber\\
   &=&\frac{f}{h(\mu,\nu)}\left[\begin{array}{ccc}
        \white{-}\sinh\mu\cos\nu&\white{..} &\cosh\mu\sin\nu \\
               - \cosh\mu\sin\nu&\white{..} &\sinh\mu\cos\nu \\
     \end{array}\right]\left[\begin{array}{c}
       \hat{x} \\
       \hat{y}\\
     \end{array}\right]
\eeqn

\noindent This can be written compactly as 

\beq
  \hat{e}_i = E_{ij}\hat{x}_j,
\eeq

\noindent where $E_{ij}$ is the elliptic rotation matrix. Its inverse is 

\beqn
   E^{-1}&=&\frac{f}{h}\left[\begin{array}{ccc}
                 \sinh\mu\cos\nu&\white{..}&       -\cosh\mu\sin\nu \\
                 \cosh\mu\sin\nu&\white{..}&\white{-}\sinh\mu\cos\nu \\
     \end{array}\right].
\eeqn

The velocity is 

\beq 
   u_i = h_i \dot{q}_i \hat{q}_i, 
\eeq

\noindent which means

\beqn
  \v{u} &=& \dot{x}\hat{x} + \dot{y}\hat{y}, \nonumber \\
        &=& h_\mu\dot\mu\hat\mu + h_\nu\dot\nu\hat\nu. 
\label{eq:uh-ell}
\eeqn

We can also get the velocity by the rotation matrix 

\beq
  u_{\he_i} = E_{ij} u_{c_j},
\eeq

i.e., 

\beqn
  u_\hmu = E_{11}u_x + E_{12} u_y,\\
  u_\hnu = E_{21}u_x + E_{22} u_y.
\eeqn

Yielding the variation of the coordinate bases

\beqn
 \dot\mu &=& fh^{-2} \left(\white{-}\sinh\mu\cos\nu \,u_x + \cosh\mu\sin\nu \,u_y\right), \\ 
 \dot\nu &=& fh^{-2} \left(-\cosh\mu\sin\nu \,u_x + \sinh\mu\cos\nu \,u_y\right).
\eeqn

\section{Vortex motion}

Consider a vortex in Cartesian coordinates

\beqn
u_x &=& -\varOmega y \chi\\
u_y&=& \white{-}\varOmega x /\chi
\eeqn

We seek to transform this into elliptic coordinates. The 
vortex motion occurs on ellipses of constant eccentricity, whereas 
the elliptic coordinate system defines confocal ellipses of different 
eccentricity. An elliptic coordinate system based on constant 
eccentricity \citep{ChangOishi10}, although matching
the flow geometry, is not orthogonal, which complicates
analysis \citep{LyraLin13}. If the streamlines matched the eccentricities of the 
confocal ellipses, the velocity would everywhere reduce to
$\dot\mu=0$ and $\dot\nu={\rm const}$. However, that is not the case, as one can verify that 
this is not divergenceless. In fact, there is only one streamline
that obeys $\dot\mu=0$ and $\dot\nu={\rm const}$, which is the
streamline of eccentricity matching the eccentricity of the
vortex. This is the particular ellipse $\mu_0$, given by $\tanh\mu_0
= \chi^{-1}$. We write the velocities as  

\beqn
  u_x &=& -\varOmega\,f\,\cosh\mu\sin\nu, \\ 
  u_y &=& \white{-}\varOmega\,f\,\sinh\mu\cos\nu.
\eeqn

We transform these into elliptical coordinates by the rotation matrix 

\beq
  u_{\he_i} = E_{ij} u_{c_j}
\eeq

\noindent yielding

\beqn
u_\hmu &=& -\Omega\frac{f^2}{2h}\left(\frac{\cosh\,2\mu-\cosh\,2\mu_0}{\sinh\,2\mu_0}\right)\sin\,2\nu, \label{eq:uhmu}\\
u_\hnu &=&  \Omega\frac{f^2}{2h}\left(\frac{\sinh\,2\mu}{\sinh\,2\mu_0}\right)(\cosh\,2\mu_0-\cos\,2\nu). \label{eq:uhnu}\\
\eeqn

The divergence is 

\beqn
  \Div{A} &=& u^{\hat\alpha}_{;\hat{\alpha}} =  u^{\hat\alpha}_{,\hat{\alpha}} + \Gamma^{\hat\alpha}_{\hat\beta\hat\alpha}u^{\hat\beta},\nonumber\\
          &=& u^{\hat\alpha}_{,\hat{\alpha}} + \frac{f^2}{2h^3}\left( u^\hmu \sin\,2\nu + u^\hnu\sinh\,2\mu\right), \\
\eeqn

or, abandoning the co-variant formulation,

\beq
  \Div{A} = \frac{1}{h^2}\left(\pderiv{h u_\hmu}{\mu} + \pderiv{h u_\hnu}{\nu}\right).
\eeq

\noindent we conclude that the flow is divergenceless. 

\eq{eq:uhmu} and \eq{eq:uhnu} may seem daunting at first, but
following the motion at the ellipse of $\mu$=$\mu_0$ simplifies it
considerably. For $\mu=\mu_0$, \eq{eq:uhmu}
cancels. For \eq{eq:uhnu}, the factor in parentheses becomes 
unity; the next term, given \eq{eq:otherh}, is $2h/f^2$. Thus, for
$\mu$=$\mu_0$, the motion is $u_\hmu$=0,
$u_\hnu$=$\varOmega\,h_0$. Comparing with \eq{eq:uh-ell} yields

\beqn
\dot{\mu} &=& 0,\\
\dot{\nu} &=& \varOmega.
\eeqn

\noindent For the particular $\mu=\mu_0$ ellipse, the motion has constant $\mu$: a closed elliptic streamline. 
The angle $\nu$ rotates uniformly. Notice that this does not mean 
that the velocity itself is uniform, since $h$ depends on $\nu$. The 
explicit dependency of $u_\nu$ on $\nu$ is 

\beqn
  u_\nu^2 &=& \frac{\varOmega^2\,f^2}{2}\left(\cosh\,2\mu - \cos\,2v\right)\nonumber\\
         &=& \varOmega^2\,f^2\left(\sinh^2\mu + \sin^2\,v\right).
\eeqn

\section{Energy conservation}

That the kinetic energy $K=u_\nu^2/2$ depends on $\nu$, a function of time, 
may seem strange at first. We show that this happens because 
the velocity change is compensated by a change in pressure ($p$), conserving
the total energy. The energy equation is 

\beq
  \frac{\partial E}{\partial t}  =
  -\Div{\left[\v{u}\left(E + p\right)\right]} +\v{F}_b\cdot\v{u} 
\eeq

\noindent where $\v{F}_b$ is a body force. The
total energy is $E=K + \varepsilon$, where 
and $\varepsilon=k_BT$ is the internal energy ($k_B$ is Boltzmann's constant
and $T$ is the temperature). In the absence of a body force and for constant temperature, this reduces 
to 

\beq
  \frac{\partial}{\partial t}\left(\frac{u^2}{2}\right) =
  -\left(\v{u}\cdot\grad\right)\left(\frac{u^2}{2} + p/\rho\right)
\eeq

\noindent therefore 

\beq
  \frac{d}{d t}\left(\frac{u^2}{2}\right) = 
  -\left(\v{u} \cdot\grad\right) p/\rho
\eeq

The enthalpy is found by Euler's equation

\beqn
 \partial_x p/\rho &=& -u_x\partial_x u_y = \varOmega^2  x\\
 \partial_y p/\rho &=& -u_y\partial_y u_x = \varOmega^2 y
\eeqn

Taking the $x$-derivative above and the $y$ below, we find
$\Laplace{p/\rho} = \varOmega^2$; therefore 

\beq
  p/\rho = \frac{1}{2}\varOmega^2\left(x^2+y^2\right)+C
\eeq

\noindent which is an intriguing result: an incompressible 
elliptical vortex produces an axis-symmetric pressure
distribution. Transforming into elliptic coordinates and eliminating
the constant 

\beq
  p/\rho = \frac{1}{2}\Omega^2f^2(\cosh^2\mu\cos^2\nu+\sin^2\mu\sin^2\nu).
\eeq

\noindent Along the $\mu_0$ streamline, the advection reduces
to the $\nu$-term

\beqn
-\left(\v{u} \cdot\grad\right) P/\rho &=& -u_\nu h^{-1}\partial_\nu p/\rho \nonumber\\
&=&\frac{1}{2}\varOmega^3f^2\sin\,2\nu,
\eeqn

\noindent whereas the time derivative of the kinetic energy is 

\beqn
  \frac{d}{d t}\left(\frac{u^2}{2}\right) &=& \Omega^2h\frac{dh}{dt}\nonumber\\
  &=&\frac{1}{2}\Omega^3f^2\sin\,2\nu.
\eeqn

That the two variations match amounts to conservation of 
energy: along the ellipse, the material slows down or speeds up 
in order to match the pressure variation. 


\section{Euler equation in elliptical coordinates}

We consider now the force balance. We use the transformations here derived to write the Euler 
equation in elliptic coordinates

\beq
  \partial_t\v{u} + \left(\v{u}\cdot\grad\right)\v{u} = -\grad{p/\rho}.
\eeq

Using covariant derivatives, this reads 

\beq
  \partial_t u^{\hat{k}} + u^{\hat{p}}\partial_{\hat{p}}u^{\hat{k}} + \Gamma^{\hat{k}}_{\hat{m}\hat{n}}u^{\hat{m}}u^{\hat{n}} = -\partial_{\hat{k}}{p/\rho}.
\eeq

For $\mu$, the correction due to the Christoffel symbols is 
\beqn
 \Gamma^\hmu_{\hatm\hatn}u^{\hatm}u^{\hatn} &=& \Gamma^{\hmu}_{\hnu\hmu},\nonumber\\ 
&=&\frac{f^2}{2h^3}\left[u^{\hnu} u^{\hmu}\,\sin\,2\nu - (u^{\hnu})^2\sinh\,2\mu\right].
\eeqn

The same procedure for $\nu$ yields

\beqn
 \Gamma^{\hnu}_{\hatm\hatn}u^{\hatm}u^{\hnu} &=& \Gamma^\hnu_{\hnu\hmu} u^\hmu u^\hnu + \Gamma^\hmu_{\hmu\hmu} (u^{\hmu})^2, \nonumber\\
&=&\frac{f^2}{2h^3}\left[u^\hnu u^\hmu\,\sinh\,2\mu - (u^\hmu)^2\,\sin\,2\nu\right].
\eeqn

Abandoning the co-variant notation

\beqn
  \partial_t\,u_\mu &=& - \left(h^{-1}u_\mu\partial_\mu +    h^{-1}u_\nu\partial_\nu\right) u_\mu -h^{-1}\partial_\mu{p/\rho} - \frac{f^2}{2h^3}\left(u_\nu u_\mu\,\sin\,2\nu - u_\nu^2\sinh\,2\mu\right),\\
  \partial_t\,u_\nu &=&  - \left(h^{-1}u_\mu\partial_\mu + h^{-1}u_\nu\partial_\nu\right) u_\nu -h^{-1}\partial_\nu{p/\rho} - \frac{f^2}{2h^3}\left(u_\nu u_\mu\,\sinh\,2\mu - u_\mu^2\,\sin\,2\nu\right).
  \eeqn


This differs from the usual equations by the presence of the extra force

\beqn
\v{F} &=& \frac{f^2}{2h^3}\left[\left(u_\nu u_\mu\,\sin\,2\nu -  u_\nu^2\sinh\,2\mu\right)\hat\mu +  \left(u_\nu u_\mu\,\sinh\,2\mu - u_\mu^2\,\sin\,2\nu\right)\hat\nu\right] 
\eeqn

Contracting this force with the velocity yields $\v{F}\cdot\v{u} = 0$, which 
shows that this force is inertial. For the vortical flow, again
following the $\mu_0$ streamline where $\dot{\mu}=0$ and $\dot{\nu}=\varOmega$, these reduce to 

\beqn
  -\frac{\varOmega^2f^2}{2}\sinh\,2\mu_0 &=& -\partial_\mu{p/\rho} \label{eq:uniform1}\\
  \left(u_\nu\partial_\nu\right) u_\nu &=& -\partial_\nu{p/\rho}
\eeqn

I.e, a constant centrifugal force that balances the  
normal pressure gradient, and inertia in the tangential direction exchanging 
kinetic energy with the pressure field. Fig~1 sketches the forces. 

\begin{figure}
  \begin{center}
    \resizebox{\textwidth}{!}{\includegraphics{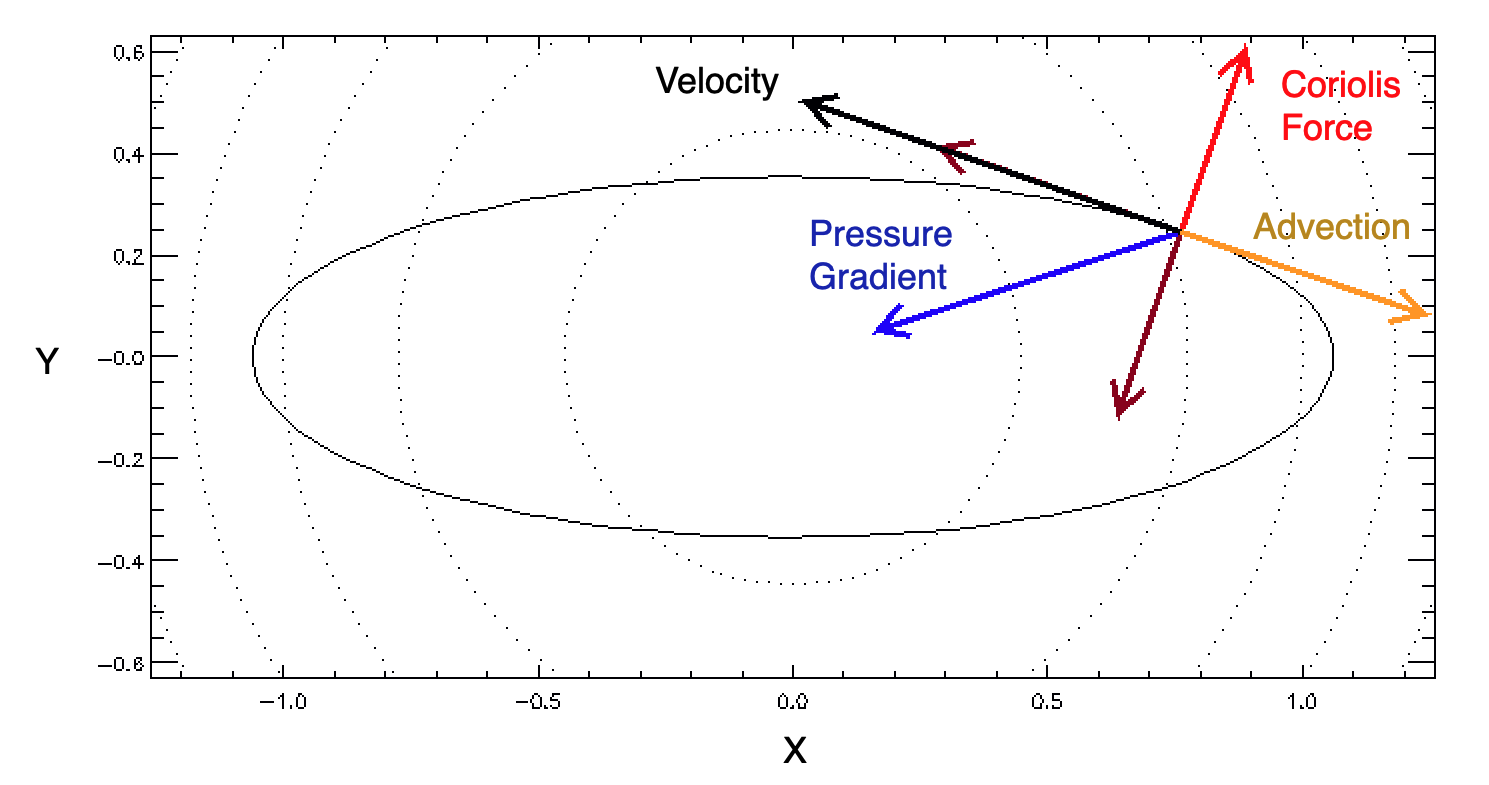}}
\end{center}
\caption{Force balance in an elliptic vortex streamline (solid
  line). Dotted circles represent the pressure contours. The
  velocity (black arrow) is tangent to the streamline, in the $\hat{\nu}$
  direction. The pressure gradient (blue arrow) is broken down in its
  $\hat{\mu}$ and $\hat{\nu}$ components (brown arrows). The $\hat{\mu}$
  component is balanced by the centrifugal force (red arrow); the
  $\hat{\nu}$ component is balanced by advection.}
\label{fig:ellipticforcebalance}
\end{figure}

\bibliographystyle{aasjournal}


\end{document}